**Title:** Unique flow features of the silent southern Boobook owl wake during flapping flight

Jonathan Lawley[1], Hadar Ben-Gida[2], Krishnan Krishnamoorthy[1], Erin E. Hackett[1], Gregory A. Kopp[3], Gareth Morgan[4], Christopher G. Guglielmo[5], Roi Gurka[1*]

[1]Department of Coastal and Marine Systems Science, Coastal Carolina University, Conway, SC, USA
[2]Faculty of Aerospace Engineering, Technion, Haifa, Israel
[3]Department of Civil and Environmental Engineering, University of Western Ontario, London, ON, Canada
[4]African Lion Safari, Cambridge, ON, Canada
[5]Department of Biology, University of Western Ontario, London, ON, Canada

Corresponding Author:
Roi Gurka
School of Coastal and Marine Systems Science
Coastal Carolina University
Conway, SC, 29579
USA
Email: rgurka@coastal.edu

# Abstract

The mechanisms associated with an owl's ability to fly silently have been the subject of scientific interest for many decades and a source of inspiration in the context of reducing noise in both flapping and non-flapping flight. Here, we characterize the near wake dynamics and associated flow structures that are produced by flying owls. The goal is to shed light on unique flow features that result from the owls' wing morphology and its motion during forward flapping flight. We study the wake of the southern boobook owl (*Ninox boobook*); a mid-sized owl, which shares the common feature of stealthy flight. Three individual owls were flown, separately, in a climatic avian wind tunnel at their comfortable speed. The velocity field in the wake was sampled using long-duration high-speed Particle Image Velocimetry (PIV) while the wings' kinematics were imaged simultaneously using high speed video. The time series of velocity maps that were acquired over several consecutive wingbeat cycles enable us to characterize the wake patterns and associate them with the various phases of the wingbeat cycle. Results reveal that the owl's wake is significantly different compared with other birds (western sandpiper, *Calidris mauri*; European starling, *Strunus vulgaris*). The near wake of the owl did not exhibit any apparent shedding of organized vortices. Instead, a more chaotic wake pattern is observed, in which the characteristic scales of vorticity (associated with turbulence) are substantially smaller in comparison to other birds. Estimating the pressure field developed in the wake depicts that the owl reduces the pressure to approximately zero. It is therefore conjectured that owls manipulate the near wake to suppress the aeroacoustic signal by controlling the size of vortices generated in its wake, which are associated with noise reduction through suppression of the pressure field.

# 1. Introduction

The silent flight of owls has been the subject of scientific interest for over a century [1,2]. Over millions of years of evolution, these species have produced many specialized configurations. Yet, most members from the order Strigiformes have the common characteristic of silent flight, which is nearly inaudible to humans and, more importantly, to their prey [3,4,5].

Graham[6] was the first to identify three unique characteristics of the owl feathers associated with silent flight; (i) the leading-edge serrations: a comb of evenly-spaced bristles along the wing leading-edge, (ii) the trailing-edge porous fringe of feathers and (iii) downy porous feathers distributed over the upper wing surface. Graham observed that the down feathers feature long hairs and barbs, which he hypothesized would muffle any rustling noise associated with the rubbing of the feathers together. He also suggested that these feathers acted as a sound absorber, which would dampen out any small vibrations near the wing, or that they worked in conjunction with the leading-edge serrations to further slow the airflow in the boundary layer of the wing. Graham [6] emphasized the importance of the leading-edge serrations as a major noise reducer. Since then, numerous studies have been performed on these three wing characteristics in attempt to isolate and identify the prime mechanisms associated with noise reduction and to relate them to the aerodynamic performance.

Kroeger et al. [7] and Anderson [8] investigated the aerodynamic role of the leading-edge serrations and found that they are responsible for turning the flow close to the wing leading-edge in a spanwise direction (toward the wingtip). Consequently, a stationary leading-edge vortex (LEV) forms [9] that delays flow separation and produces non-linear lift on the outer half of the owl's wing. Soderman [10] adapted owl-like leading-edge serrations in low-speed rotors and measured the acoustic effects at different operating conditions. It was found that leading-edge serrations were effective in reducing high frequency noise, yet the rotor performance was relatively unaffected. Rao et al. [11] used a bio-inspired model to investigate the noise reduction from a wing with leading-edge serrations. They observed a significant noise reduction at high frequencies as compared to a smooth leading-edge model. Winzen et al. [12] studied the role of leading-edge serration on the aerodynamic performance of an owl using a wing model tested in a wind tunnel. They showed that leading-edge serration serves as a flow stabilizing device, yet at the expense of a significant reduction in its aerodynamic performance. Similarly, a recent study by Geyer et al. [13] reported an owl's wing model with critical leading-edge serrations yielded a small increase in lift and some reduction in aerodynamic noise during gliding flight.

The flow past the leading-edge over the wing surface may be separated depending on the wing's angle of attack and its camber. It was suggested that the down feathers play a role in delaying separation as compared to smooth surfaces. Klan et al. [14,15] studied the flow above barn-owl wing models using flow visualization and PIV. They found that a velvet surface (mimicking the down feathers) reduced the size of the separation bubble and consequently shifted the point of transition upstream over the wing. Winzen et al. [16] studied the flow field around an owl model wing focusing on the influence of the down feathers, which was mimicked by an artificial surface that covered the wing model. They showed that the separation bubble over the wing was reduced and, in some cases, eliminated. They also demonstrated a decay of the vortical structures along the wing. Winzen et al. [17,18] subsequently showed that the down feathers as well as the flexibility of the wing stabilize the flow field at low Reynolds numbers, enabling the owl to fly more slowly.

Once the flow passes the wing, it is shed from the trailing-edge towards the wake region. Bachman et al. [19] studied the fringes at the trailing-edge and demonstrated that these merge into so-called "neighboring feather vanes" which results in a smooth lower wing surface and, thus, reduces sharp and noisy edges. Jaworski and Peake [20] suggested that the trailing-edge is the primary source of noise for aerodynamic structures in the flow. Using theoretical modeling, they showed that some morphological features in the trailing-edge region cause noise degradation. Geyer et al. [21,22], who examined the noise generation of prepared birds' wings of different species, confirmed that the silent flight of owls (i.e.: Tawny and Barn owls) is due to their special wing and feather adaptations.

Since different owl species range in wing size from ten centimeters to close to a meter, their typical gliding Reynolds number range is within the intermediate range of the order of $10^5$; therefore, low Reynolds number theories [23] cannot predict and accommodate the unique flight pattern and noise reduction exhibited by owls. Measurements by Kroeger et al. [7] indicated some differences between noise emitted during gliding and flapping flight; whilst Jones et al. [24], and later on Platzer et al. [25], presented the advantages of flapping flight in producing thrust and lift. It appears that owls stealth mechanisms [26] function regardless of their flight mode: gliding or flapping. Owls feature highly maneuverable low-speed gliding flight capabilities [27], yet their aerodynamic performance (lift-drag ratio) observe to be relatively low [7].

While the owl's wing morphology has been studied extensively over the years [28,29] limited work has been focused on the aerodynamics of owls and the interaction between their unique wing morphology and wake flow dynamics. Doster et al. [30] flew a trained barn owl (*Tyto alba*) in a wind tunnel towards a target, where they performed Stereo-PIV measurements around the owl during flight. They showed that a complex vortex flow system was developed at the wake during flapping flight, yet, they have not associated these complex dynamics with noise reduction. The presented work focuses on the interaction between a freely flying owl and the surrounding fluid, where we seek to shed light on the governing flow mechanisms associated with silent flight. The results presented herein shows how owls control the wake flow through modulations of the flow scales and suppression of turbulence.

## 2. Methods

Owls were flown in an avian wind tunnel where optical flow measurement techniques were employed to study the near wake flow as well as the owls' kinematics during flapping flight.

*Birds*

Three Boobook owls (*Ninox boobook*) were tested; two males and one female. This owl is a medium sized nocturnal owl with stealth capabilities. The owls were brought from the "African Lion Safari" in Cambridge, ON, Canada under animal protocol number BOP-15-CS and protocol 2010-2016 from the University of Western Ontario Animal Care committee. Morphological parameters of the owls, as well as the non-dimensional flow numbers associated with the performed experiments are summarized in Table 1. The owls adjusted quickly to the wind tunnel and performed successful flights over several days. A set of optoisolators operated by six infrared transceivers were integrated into the PIV system (upstream from the laser sheet location) in order to prevent direct contact between the bird and the laser sheet [31]. The optoisolators triggered the laser only when the owl was flying upstream from the PIV field of view. The isolation from the laser sheet and triggering system ensure the safety of the birds.

*Wind tunnel*

The owls were flown in the hypobaric climatic wind tunnel at the Advanced Facility for Avian Research (AFAR) at the University of Western Ontario (see Kirchhefer et al. [31] for more details). The wind tunnel is a closed loop type with a glass octagonal test section of 2 m length, 1.5 m width and 1 m height preceded by a 2.5:1 contraction. The turbulence intensity at the test section was smaller than 0.3% with a velocity profile uniformity of 0.5%. Speed, pressure, temperature and humidity can be controlled to generate various flight conditions at different altitudes. The bird is introduced into the test section through a 0.5 m open jet section located between the downstream end of the test section and the diffuser. The flight conditions for all the owls were at atmospheric pressure, a temperature of 15°C and a relative humidity of 80%.

*Particle image velocimetry (PIV)*

A long-duration time-resolved PIV system [32] was employed to measure the near wake field behind the owls' wings during flight. The PIV system consisted of a 80W double-head diode-pumped Q-switched Nd:YLF laser operating at a wavelength of 527 nm and two CMOS cameras (Photron FASTCAM-1024PCI) with spatial resolution of 1024x1024 pixel$^2$ operating at a rate of 1000 Hz and a 10 bit dynamic range. The PIV system is capable of continuously acquiring image pairs at 500 Hz using two cameras for 20 min. Olive oil particles, 1 μm in size were introduced into the wind tunnel using two Laskin nozzles [33] from the downstream end of the test section; thus, it did not cause a disturbance to the flow or to the bird.

One of the CMOS cameras was used for the PIV, while the other CMOS camera was used for measuring the wingbeat kinematics simultaneously with the PIV. The PIV camera's field of view (FOV) was 13x13 cm$^2$ corresponding to $1c$x$1c$, where $c$ is the owl mean chord length (an average of the three owls' chord lengths) and the kinematic camera's field of view was 52x52 cm$^2$ corresponding to $4c$x$4c$, as shown in Figure 1. The near wake flow field was sampled in the streamwise-normal plane with temporal resolution of 500 Hz and the distance from the owl's trailing-edge to the FOV varied between 1.4 to 4 chord lengths. Multiple experiments were conducted to sample the near wake flow field at different locations along the wing span of the owl, which enabled us to comparatively analyze the wake characteristics developed at different wing sections along the span at each wingbeat cycle. Table 2 summarizes the collected PIV data-sets obtained during the experiments for the three owls. The velocity fields were computed using OpenPIV [32] with 32x32 pixel$^2$ interrogation windows and 50% overlap, yielding a spatial resolution of 64 vectors per average chord, equal to 1.8 vectors per mm. The coordinate system used is a right-handed Cartesian system, where $x$, $y$, $z$ corresponds to the streamwise, normal and spanwise directions. $x$ is directed downstream, $y$ is directed upwards and $z$ is determined according to the right-hand rule. The streamwise and normal velocity components are denoted by $u$ and $v$, respectively.

*Kinematics*

Wingbeat kinematics were recorded using one of the high-speed CMOS cameras as described in the previous section. From the acquired images, we calculated the wingbeat frequency, wingbeat amplitude, and angle of attack. The kinematic images were synchronized with the PIV images in order to provide a direct relationship between the wake formed by the wing motion and its kinematics.

The wake locations with respect to the trailing-edge of the wing during flight was determined from both captured images and the distance between them as depicted in Figure 1. To determine the location of the light sheet along the bird's wing or body, a 30 Hz CCD camera with 1600x1200 pixel$^2$ resolution was mounted downstream of the test section pointing towards the location where the bird would trigger the laser and taking spanwise-normal plane images. A spatial calibration was performed before the experiment. Once synchronized, spanwise positions were assigned to the wake data acquired at 500 Hz based on interpolation from the simultaneously recorded spanwise positions. These images enabled us to identify the location of the light sheet relative to the wing during the experiments and determine its location in respect to the body center. This information allows us to couple the wake flow features with the wing morphology: wingtip, primary and secondary remiges and close to the body (at the root).

*Experiments*

Table 2 summarizes the collected PIV data sets obtained during the experiments for the three owls that were analyzed in this study. Only successful owl flights were recorded. Successful flight refers to experiments where the owl triggered the laser and the PIV system acquired images of the flow field simultaneously with the wing's kinematics. The data presented herein corresponds to sets ('scenes') where the owl did not accelerate or decelerate and maintained altitude during flapping mode. A total of 9 scenes are presented herein, where each scene consists of hundreds of vector maps that corresponds to multiple consecutive wingbeat cycles during free flight. This large set of data enabled us to, statistically, characterize the near wake flow field, and its interaction with the owl's wing. The PIV measurements were taken behind the owl's wing where a wake was present, and the wing motion was clearly identified in the kinematic images.

An error analysis based on the root sum of squares method was applied to the velocity data and the wing kinematics, following Gurka et al. [34]. The errors were estimated as: 2.5% for the instantaneous velocity values, 10% for the instantaneous vorticity and 4% for the circulation, which was calculated, based on the vorticity field [35]. The error introduced in the kinematic analysis resulted from the spatial resolution of the image and the lens distortion leading to an estimated error of 5% in the wing displacements.

## 3. Results

*3.1 Owl wing kinematics*

We use a similar approach as Gurka et al. [34], which followed guidelines suggested by Wies-Fogh [23], where the wingbeat cycle was divided into four distinct phases: upstroke (US), transition from upstroke to downstroke (USDS), downstroke (DS) and transition from downstroke to upstroke (DSUS). Figure 2 presents images in sequential order from right to left as the owl flies through one full wingbeat cycle during forward flight (upstream, against the wind). We extracted the wingtip motion of the owls using freeware motion analysis software, Kinovea (https://www.kinovea.org). A point close to the wingtip was tracked for all the three owls over the continuous wingbeat cycles. The tracking identification for the kinematics analysis varied from scene to scene, but always located at the tip of the third primary wing feather. The number of wingbeat cycles was calculated by normalizing the total evaluation time with the wingbeat period. We estimated that for the various flight durations, the average frequency was about 6 Hz (see figure 3). Therefore, the corresponding Strouhal Number for owls #2 and #3 is 0.37 and for owl #1 is 0.35. Herein, the Strouhal Number is defined as $St = fA/U_\infty$, where $f$ is the flapping frequency, $U_\infty$ is the speed of flight (wind tunnel speed), and $A$ is the peak-to-peak cross stream amplitude of the motion [36]. A semi-sinusoidal pattern is observed, covering the upstroke, downstroke and transition phases over almost two consecutive wingbeat cycles. The axes presented are normalized by the chord length. The trend is similar for various data sets covering flights of the three owls. The solid vertical lines define the transitions between the wingbeat phases whereas the dashed line illustrates the point during the downstroke phase where the angle of attack was calculated.

The estimation of the wing's angle of attack (γ), as depicted in table 3, is taken from the pitch of the chord line at the root of the wing and velocity of the wing's root relative to the air[31]. Determining the angle of attack of a wing that constantly flaps and twists is somewhat challenging; therefore, we chose a point when the wing tip was parallel to the body. Assuming constant wing twist, the angle of attack is defined as the angle between the chord relative to the air stream and is computed using information from the flow, body and chord at the root (see figure 4). For a flapping wing, this angle is comprised of i) the angle, α which is estimated by $\tan^{-1}(w/u)$, where $w$ is the velocity of the leading-edge at the root (obtained from the kinematic images) and $u$ is the streamwise velocity (obtained from the PIV) and ii) the angle between the root chord and the body, β. By taking the difference between the latter angle (ii) and the former angle (i) an estimate of the angle γ, for the root chord of the flapping wing can be made. It is noteworthy that the angle of attack is calculated at relatively the same phase for each wingbeat of the scene: when the leading-edge is relatively parallel to the shoulder. Assuming the wing has a constant twist throughout the downstroke, the wing's angle of attack, at the root, varied between 5º to 18º. This range, including the relatively high maximum angle of attack, is common in other birds as well [37]. These high angles of attack allow owls to fly more slowly while still being able to generate lift [38] through the formation of LEV [8].

*3.2 Near wake fluid dynamics*

The near wake of the owl may provide insight into how wing morphology combined with wing kinematics enable the owl to fly silently. At such stealthy flight mode, the owl suppresses the aeroacoustic noise, partially as a result from the wake flow dynamics. The results presented in this section demonstrate that the owl wakes differ from those of other birds, which are more similar to each other. In order to qualitatively assess the wake evolution resulting from the owl flapping flight, we follow the same procedure as originally suggested by Spedding et al. [39] and utilized later for other passerines and shorebirds [31,34]. The wake evolution in time, which can be transformed into the evolution in space enables one to observe how the vortical patterns in the wake region provide a unique signature of a bird's flight. The wake reconstruction procedure we used is described in detail in Gurka et al. [34]. It is noted that throughout the presented wingbeat cycles, the owls' position did not change much relative to the measurement plane. Therefore, Taylor's hypothesis [40] is applied, following the assumption that the flow remains relatively unchanged as it passes through the measurement plane. The utilization of the long-duration time-resolved PIV system enabled the reconstruction of the wake evolving behind the wings. The owls flew from right to left (figure 2); therefore, the downstream distance is measured as positive chord lengths. What appears as downstream essentially happened earlier while what appears as upstream happened later. Each wingbeat cycle corresponds to 5 to 8 cord lengths for the various wakes analyzed. Each individual scene analysis corresponds to 0.5-2 wingbeat cycles; thus, we can analyze the flow field behind the owl continuously and identify trends within the flow patterns.

Figure 5 (top) demonstrates the evolution of the near wakes behind the freely flying owls. The color contours correspond to the spanwise vorticity and the velocity vectors are superimposed on the contour maps. The regions of positive and negative vorticity are marked in red and blue, respectively. The lower subplots in figure 5 depict the motion of the wing which is coupled with the flow in the wake above. The spanwise vorticity was normalized by the ratio between the chord length $c$ and the free-stream velocity, $U_\infty$. All the wakes presented in figure 5, have been calculated based on the same threshold of the normalized vorticity values (-1 to +1). The axes are scaled based on the specific owl's chord length, measured at the semi-span location (between the primary and secondary remiges). It appears that the shedding of vortices from the wing are somewhat lacking coherence or consistency where one would expect to observe some sort of shedding behavior; organized or non-organized from a propulsive wake [31,34,39,41]. The vorticity patterns in the wake appear disorganized, as shown in multiple sets of the data for the three owls investigated (see figure 5). Figure 5a depicts the wake reconstruction from data taken in experiment #9. The wake presented corresponds to the flow formed above and below the wing section, located between the primary and secondary remiges. Additional datasets acquired at the same location are shown in the supplementary material for experiments #1, 2, 4 and #7 (see ESM1). Experiment #8, presented in figure 5b, corresponds to the wake formed at the outer region of the wing; the furthest location in the primary remiges, where a tip vortex is present. The tip vortex appears as a concentrated spanwise vorticity region, almost circular in its geometrical shape marked with strong positive and negative vorticity values preceded by weak shedding that occurs over the entire wingbeat cycle. Additional datasets acquired at the same location are shown in the supplementary material for experiments #5 and #7 (see ESM1). It is noteworthy that experiment #7 had mixed wake flow patterns, which may indicate that the owl was moving in the spanwise direction during flight. Figure 5c depicts the wake behind the middle point of the primary remiges (Experiment #3). Experiment #6, which presents the wake behind the secondary remiges close to the root, is depicted in the supplementary material (ESM1). In general, one may conclude that the common feature for all the wakes examined is that the concentrated regions of spanwise vorticity are small, suggesting that small scales dominate the wake flow. This qualitative examination of the reconstructed wakes shows a different topography of the vorticity field in comparison to other birds that were tested in the same facility [34]. The near wake flow of passerines such as European starling (*Sturnus vulgaris*) and America robin (*Turdus migratorius*) and a shorebird (western sandpiper, *Calidris mauri*) exhibit an organized wake where shedding is observed, although these birds are of different size and flight behavior. This discrepancy suggests the owl is generating a wake inherently different when compared with these other birds. In order to quantitatively characterize these differences, we have performed a topographical and flow-scale analysis of the reconstructed wakes of an owl, starling, and sandpiper. The results of the analysis of each bird are compared and contrasted to enable a measure of the distinction of the wakes.

For the topographical analysis, a quantitative comparison between wake composites of the three distinct birds is performed through a so-called "blob" analysis. The motivation of this analysis is to characterize the dominant spatial scales in the reconstructed wake. The vorticity contours are presented in figure 5a here, and in figures 4 and 5 in Gurka et al. [34]. All the contours plotted employ a threshold of -1 and +1 for the normalized vorticity. The blob analysis essentially calculates the area of the concentrated vorticity regions ($\omega_z c/U_\infty < -1$ and $\omega_z c/U_\infty > 1$ and computes a histogram of these areas. A more detailed description of the procedure can be found in supplementary material ESM2. For brevity, the analysis transforms vorticity contour images to greyscale images and removes the background. The greyscale image is filtered and then converted into a binary image. The binary image is evaluated for interconnectivity of non-zero pixels using a connectivity of 8 nodes. The sums of the connected pixels are used to compute an area along with the image calibration from the PIV measurements. The histogram of these areas is computed and power density functions (PDF) are fit to the histograms. Figure 6 depicts this histogram for experiment #9 for the owl, along with the corresponding histograms for the starling and sandpiper wakes. The two subplots represent the histograms for the identified areas with positive vorticity (top figure) and the areas associated with negative vorticity (bottom figure). The histogram distribution for the starling and sandpiper appear to be similar, spanning a range of areas ($0.2 \times 10^{-4}$ - $2.0 \times 10^{-4}$ $m^2$), with a large standard deviation. In comparison, the owl histogram is more narrowly distributed with a lower mean area than the other birds. These results are consistent over the range of experiments presented here (see ESM1). The measured mean and standard deviation of the blob analysis histograms of the owl are smaller than the other two birds and provided in table S1 in supplement ESM2. These results demonstrate that regions of large magnitude vorticity ($\omega_z c/U_\infty > |1|$) in the wake of the owl tend to be smaller than that of other birds relative to their chord size. The limited large-scale motion in the owl's wake suggests that large-scale motion may be suppressed/damped or not generated at all. This result is also consistent with the qualitative comparison of the wakes (figure 5a), where the wake reconstruction of the owl appears to exhibit a disorganized shedding compared to the starling and sandpiper.

For the flow scale analysis, we estimated a characteristic flow scale using auto-correlation functions applied to the data in the near wake region (see details of the birds investigated in table 4). The flow in the wake is unsteady and turbulent due to the intermediate Reynolds number. Smith et al. [42] showed a linear relationship between concentrated regions of high vorticity and turbulent flow scales in homogenous turbulence [42]. The characteristic turbulent scale is known as the integral length scale which is calculated from the auto-correlation function of the fluctuating velocity field in respect to a prescribed direction [43]. The wake developed behind the freely flying owl is unsteady and one cannot use the classical Reynolds decomposition in order to extract the velocity fluctuations. Therefore, to estimate the fluctuating part of the flow, we applied a local Galilean decomposition; i.e.: $u'=u-u_{avg}$, $v'=v-v_{avg}$ where $u_{avg}$ and $u_{avg}$ are the spatially averaged (over the PIV FOV) velocities of the velocity components examined over the direction of the correlation. A similar technique was applied to PIV data in shear flows to remove the convection velocity [44]. The analysis does not attempt to estimate turbulent properties based on this decomposition; but rather is used to calculate the auto-correlation values for a fluctuating portion of the velocity field, which presumably, is associated with turbulence. Both longitudinal and transverse scales were calculated for the two velocity components and compared between the three birds, as shown in table 5. The longitudinal scale corresponds to the result obtained from correlating the velocity component along the same direction and calculating the area under the normalized correlation curve. Correlating the velocity components along the normal direction yields the transverse scales. The scales presented are averaged over each vector map and then over time: $L_{11}$ is the longitudinal length scale for the streamwise velocity ($u$) in the streamwise direction ($x$) and $L_{22}$ is the longitudinal length scale for the normal velocity ($v$) in the normal direction ($y$). The transverse scales ($L_{12}$ and $L_{21}$) correspond to the flow scales based on the streamwise velocity ($u$) and the normal velocity (*v*) along directions normal to each velocity component, respectively. Because the wake flow results from the wingbeat motion, the flow scales were estimated for two phases during the wingbeat cycle: the transition from downstroke to upstroke (DSUS) and the transition from upstroke to downstroke (USDS). The computed scales are normalized by the respective chord lengths (see table 5). The flow scales of the owl's wake are smaller by an order of magnitude with respect to the other two birds' wakes. The flow scales of the wakes behind the starling and the sandpiper have similar magnitude. The flow scales do not seem to be dependent on the wingbeat phases. In addition, these flow scales are substantially smaller than the wing chord length and presumably are governed mainly by vorticity and/or strain. The smaller dominant flow scale found for the owl in comparison to the other birds is consistent with the results of the topographical analysis.

The results of the flow scales and topographical analysis quantitatively demonstrate how the owl's wake is fundamentally different than the two other birds. To further explore the observed scale reduction, we examine the velocity gradient tensors; mainly the vorticity and strain in the wake region. The pressure Hessian is a key quantity that controls vortex stretching through interactions associated with the pressure term [45,46] in the momentum equations for fluids. The pressure Hessian is computed by applying a divergence to the incompressible Navier-Stokes equations:

$$\nabla^2 p = \rho \left(\frac{1}{2}\omega^2 - s^2\right) \tag{1}$$

where $\omega^2 = \omega_i\omega_i$, $s^2 = s_{ij}s_{ij}$, $\omega_i$ is the vorticity vector, $s_{ij}$ is the strain rate tensor, and $\rho$ is fluid density. Note that because the data is comprised from two-dimensional flow field measurements, we can only estimate the corresponding terms contributing to the pressure

Hessian. We account for the spanwise vorticity component and the rate of strains in the streamwise and normal directions. Whilst this term provides an insight to the relation between vorticity and strain, it also provides an indirect estimate of the pressure field developed within the flow. The noise generated during flight is comprised from sound that stems from a mechanical wave propagating through a medium, and from the aerodynamic noise (aeroacoustic) that is generated though the fluid motion and its eventual interaction with a solid surface. Because one of the fundamental differences between owls and other birds is its suppression of noise and sounds association with pressure, it seems appropriate to compare the pressure Hessian at the near wake region between the other birds and the owl. Figure 7 presents the histogram distribution of the right-hand side of eq. 1 for the three birds, using the same data as the topographical and flow analyses. The right-hand side term of eq. 1 is calculated at the near wake region behind the wings, for all three birds in flapping mode about the wing mid-section. Therefore, it is plausible to assume that the locations of the measurements in respect to the wings in the spanwise and streamwise directions were similar and the phenomena observed occurred roughly at the same flow configuration. The blue colored histogram distribution in figure 7 corresponds to dataset obtained from experiment #9 (owl #1), which has a mean value near zero with a tail ranging up to -0.25 kg/m$^3$sec$^2$. In contrast, for the starling, it ranges from 0 to -2.5 kg/m$^3$sec$^2$ (red histogram) and the sandpiper ranges from 0 to -0.5 kg/m$^3$sec$^2$ (green histogram); both with non-zero mean values. The histogram distribution for the starling is flatter and spans a range of values and is similar to a normal distribution whilst the other two birds have a more skewed distribution, similar to log-normal distributions. The difference between the distributions may be attributed to the wake flow patterns, which appear to be meandering for the owl and the sandpiper and less meandering for the starling, for the data cases studied herein. It is noteworthy that the same calculation was performed for the other owls' datasets and all had similar distributions to the one presented in figure 7 (blue color histogram) with a similar range of values (see figure S2 in ESM1). Calculating 95% confidence intervals indicates that there is no overlap in the mean gradient pressure distributions among the owl ($m \pm$ SD = -0.018 ± 0.032, 95% CI -0.022 to -0.013), sandpiper ($m \pm$ SD = -0.14 ± 0.16, 95% CI -0.17 to -0.12) and the starling ($m \pm$ SD = -1.14 ± 0.43, 95% CI -1.21 to -1.08), where $m$ is the mean, SD is the standard deviation and CI is the confidence interval. These statistics show that the distribution of the pressure gradient of the owl's wake is closer to zero (at least an order of magnitude smaller compared to the other birds) with little variation. This result indicates that the wake dynamics behind the owl are fundamentally different in comparison to the other two birds - consistent with the results of the other analyses. For the starling and sandpiper, the histogram mean values (right side of equation (1)) are negative, on average. Here, $\omega^2$ corresponds to enstrophy and $s^2$ is proportional to dissipation; thus, these results imply that dissipation is more dominant in the starling and sandpiper wakes behind the wing mid-span location, relative to the enstrophy (based on the relations between them in equation (1)). Yet, for the owl, the enstrophy is approximately double the dissipation as both terms counter each other to yield values in the wake that are close to zero. Therefore, we conjecture that the owl, using its unique wing morphology, generates more vorticity than strain, which essentially is achieved by generating more small scales while destroying, or not generating, large scales in its wake. Because pressure is associated with aeroacoustic noise, we can assume that this suppression of large scales and increased production of small scales is associated with noise suppression.

## 4. Discussion

The near wake flow dynamics of an owl feature unique characteristics that we suggest are associated with its ability to fly silently. Our findings demonstrate significant differences between the wake of an owl and the wake of two other birds, starling and sandpiper. Importantly, these other two birds generate wakes that are similar to each other, despite being different species. In comparison to these other two birds, the wake of the owls is verified to be quantitatively different in terms of the scales of the flow. This result is confirmed by two different methods of estimation of the flow scales. The topographical analysis demonstrated that the owl's wake is qualitatively more disorganized (no street is apparent) at the mid-span location of the wing and contain smaller areas of large magnitude of vorticity ($|\omega_z c/U_\infty|>1$). This result is consistent with the notion that owls have relatively poor aerodynamic performance [7]. An aerodynamic body is expected to generate an organized street at the wake, which indicates relatively low drag conditions. Yet, such organized structures are absent in the owl's wake. Furthermore, the flow scale analysis, which estimates the decorrelation scale of the flow patterns in the wake, indicates a smaller scale for the owl compared to the other two birds, consistent with the topographical analysis. The apparent absence of large flow scales may suggest that the turbulence production activity associated with these scales is somewhat limited [43,46]. The dominance of small scales in the wake region also indicates an increase in the turbulence dissipation rate and vorticity. Together, this implies that over the wingbeat cycle, there is imbalance between the production and dissipation of turbulence energy.

The aspects of the flow that result in the different distribution of scales, as well as its potential relationship with the noise suppression of an owl's flight, was examined via the distribution of the two-dimensional pressure Hessian in the wake. The distribution for the owl shows that the pressure Hessian term has a mean near zero with a narrow distribution compared with the other two birds. The larger enstrophy in the owl's wake relative to the dissipation ($s^2$) further supports the results of smaller flow scales in the wake and/or suppression of larger scales. This result implies that in the owl's wake the strain and vorticity fields interact with each other differently in comparison to the other two birds because the pressure Hessian describes the non-local interaction between vorticity and strain [47]. Tsinober [46] suggested that when a flow field has a zero pressure Hessian, then the flow must be non-turbulent, or in other words, that nonlocality due to pressure is essential for (self-)sustaining turbulence. Therefore, the zero distribution suggests that turbulence is suppressed through distractive local interaction between vorticity and strain. The small mean pressure Hessian could be related to the noise suppression because noise and pressure are related, and we conjecture that the suppression of aerodynamic noise occurs through modulation of the flow scales in the wake.

Therefore, we suggest that most of the owl's wake has either i) experienced a significant degradation of the turbulence level, or alternatively ii) has a strong three-dimensional motion in the spanwise direction (not measured) such that the wake behind the primary remiges is weakened relative to the tip region. For the latter case it may be that these wake dynamics resulting from flow patterns formed above the wing section being shifted in the spanwise direction towards the wing tip such that the majority of momentum is transferred from the streamwise to the spanwise direction (as suggested originally by Kroeger et al. [7]). This shift would minimize the wake activity behind the majority of the wing by shifting all the momentum towards the tip region. This explanation is also supported by the good maneuverability capabilities of an owl in exchange for generating relatively high drag during its flight, which can be caused by the aforementioned shifting required to suppress the aeroacoustic signature. For the former (turbulence suppression), these peculiar flow dynamics may be the result of the unique morphological structure of the leading-edge serrations, down feathers, and trailing-edge fringes [6] of the owl's wing, which in consequence minimize the aeroacoustic signature [48] through the control of turbulence. Once the flow interacts with the wing at the leading edge, the serrations funnel the flow and presumably shift some of the momentum towards the tip; then the flow passes over the wing through the down feathers, which prevents separation over the wing; thus, maintaining lift and reducing friction [4]. Subsequently, the flow passes towards the trailing edge. At the trailing edge, the fringes, which are unconnected barb ends; some oriented in the streamwise direction and some oriented to overlap with the neighborhood feathers [19], generate additional mixing due to their non-structured configuration. This process causes the length scales of the flow to decrease dramatically, suppressing the large scales while producing more mixing and forming more small scales, which corresponds to additional generation of vorticity. This additional vorticity and/or suppression of larger flow length scales lead to a decreased pressure gradient field that is correlated with the aerodynamic noise. One may conclude that owls reduce noise by altering the scales of the flow. Further research into how the various morphological features of the wings modify flow scales to balance the strain and vorticity fields in such a way that the pressure gradient field is minimized should be pursued.

## Ethics

The owls were brought from the "African Lion Safari" in Cambridge, ON, Canada under animal protocol number BOP-15-CS and protocol 2010-2016 from the University of Western Ontario Animal Care committee.

## Data, code and materials

All relevant data are given within the paper.

## Competing interests

We have no competing interests.

## Authors' contributions

HBG and RG was responsible for the design of the experimental setup and performed the experiments and analyzed the data. RG, JL and EEH prepared the manuscript. JL, HBG and KK performed the data analysis and edited the manuscript. GAK assisted in the setup design and edited the manuscript. CG assisted with study design and data collection and edited the manuscript. GM was responsible for the capture, care and training of the experimental birds.

## Acknowledgment

We would like to thank Prof. Arkady Tsinober for his insightful suggestions and discussion on animal-turbulence interaction. We thank Prof. Roi Holzman for reviewing the manuscript.

# References

1. Mascha E. 1904 The structure of wing-feathers, Smithsonian Institution (originally published in 1904; *Zeitschrift für wissenschaftliche Zoologie* **77**, 606-651.

2. Dubois AD. 1924 A nuptial song-flight of the short-eared owl. *The Auk 41*, 260-263.

3. Neuhaus W, Bretting H, Schweizer B. 1973 Morphologische und funktionelle intersuchungen uber den lautlosen flug der Eule (Strix aluco) im vergleich zum flug der Ente (Anas plathyrhynchos). *Biologisches Zentralblatt* **92**, 495-512.

4. Bachmann T, Klän S, Baumgartner W, Klaas M, Schröder W, Wagner H. 2007 Morphometric characterization of wing feathers of the barn owl Tyto alba pratincola and the pigeon Columba livia. *Frontiers in Zoology* **4**, 23.

5. Sarradj E, Fritzsche C, Geyer T. 2011 Silent owl flight: bird flyover noise measurements. *AIAA J.* **49**, 769-779.

6. Graham RR. 1934 The silent flight of owls. *J. Royal Aero. Soc.* **38**, 837-843.

7. Kroeger RA, Gruschka HD, Helvey TC. 1972 Low speed aerodynamics for ultra-quiet flight. *Technical Report AFFDL TR-71-75*, The University of Tennessee Space Institute, Tullahoma, TN.

8. Anderson GW, 1973 An experimental investigation of a high lift device on the owl wing. *Technical Report No. GAM/AE/73-6*, Air Force Institute of Technology Wright-Patterson AFB OH School of Engineering.

9. Lowson MV, Riley AJ. 1995 Vortex breakdown control by delta wing geometry. *J. Aircraft* **32**, 832-838.

10. Soderman PT. 1973 Leading edge serrations which reduce the noise of low-speed rotors, *Technical Report D-7371*, NASA.

11. Rao C, Ikeda T, Nakata T, Liu H. 2017 Owl-inspired leading-edge serrations play a crucial role in aerodynamic force production and sound suppression. *Bioinsp. Biomim.* **12**(4), 046008.

12. Winzen A, Roidl B, Klän S, Klaas M, Schröder W. 2014 Particle-Image Velocimetry and force measurements of leading-edge serrations on owl-based wing models. *J. Bionic Eng.* **11**, 423–438.

13. Geyer TF, Claus VT, Hall PM, Sarradj E. 2017 Silent owl flight: The effect of the leading edge comb. *Int. J. Aeroacoustics* **16**(3), 115-134.

14. Klän S, Bachmann T, Klaas M, Wagner H, Schröder W. 2009 Experimental analysis of the flow over a novel owl based airfoil. *Exp. Fluids* **46**, 975-989.

15. Klän S, Burgmann S, Bachmann T, Klaas M, Wagner H, Schröder W. 2012 Surface structure and dimensional effects on the aerodynamics of an owl-based wing model. *Euro. J. Mech.-B/Fluids* **33**, 58-73.

16. Winzen A, Klass M, Schröder W. 2013 High-speed PIV measurements of the near-wall field over hairy surfaces. *Exp. Fluids* **54**, 1-14.

17. Winzen A, Klaas M, Schröder W. 2015 High-speed Particle Image Velocimetry and force measurements of bio-inspired surfaces. *J. Aircraft* **52**(2), 471-485.

18. Winzen A, Roidl B, Schröder S. 2015 Particle-Image Velocimetry investigation of the fluid-structure interaction mechanisms of a natural owl wing. *Bioinsp. Biomim.* **10**(5), 056009.

19. Bachmann T, Wagner H, Tropea C. 2012 Inner vane fringes of barn owl feathers reconsidered: morphometric data and functional aspects. *J. Anatomy* **221**, 1-8.

20. Jaworski JW, Peake N. 2013 Aerodynamic noise from a poroelastic edge with implications for the silent flight of owls. *J. Fluid Mech.* **723**, 456-479.

21. Geyer TF, Sarradj E, Fritzsche C. 2009 Silent owl flight: experiments in the aeroacoustic wind tunnel. *NAG/DAGA International Conference on Acoustics 2009*, 734-736.

22. Geyer TF, Sarradj E, Fritzsche C. 2012 Silent owl flight: Acoustic wind tunnel measurements on prepared wings. *18$^{th}$ AIAA Aeroacoustics Conference* (Colorado Springs, CO.

23. Weis-Fogh T. 1973 Quick estimates of flight fitness in hovering animals, including novel mechanisms for lift production. *J. Exp. Bio.* **59**, 169-230.

24. Jones KD, Lund TC, Platzer MF. 2001 Experimental and computational investigation of flapping wing propulsion for micro air vehicles. *Prog. Astro. Aeronautics* **195**, 307-339.

25. Platzer MF, Jones KD, Young J, Lai JCS. 2008 Flapping wing aerodynamics: progress and challenges. *AIAA J.* **46**, 2136-2149.

26. Thorpe WH, Griffin DR. 1962 Lack of ultrasonic components in the flight noise of owls, *Nature* **193**(4815), 594-595.

27. Johnsgard PA. 1988 North American owls: biology and natural history, Smithsonian Institution Press.

28. Weger M, Wagner H. 2017 Distribution of the characteristics of barbs and barbules on Barn owl wing feathers. *J. Anatomy* **230**(5), 734-742.

29. Wagner H, Weger M, Klaas M, Schröder W. 2017 Features of owl wings that promote silent flight, *The Royal Soc.: Interface Focus* **7**(1), 20160078.

30. Doster T, Wolf T., Konrath R. 2014 Combined flow and shape measurements of the flapping flight of freely flying barn owls. *New Results in Numerical and Experimental Fluid Mechanics IX* 661-669.

31. Kirchhefer AJ, Kopp GA, Gurka R. 2013 The near wake of a freely flying European starling. *Phys. Fluids* **25**, 051902.

32. Taylor ZJ, Gurka R, Kopp GA, Liberzon A. 2010 Long-duration time-resolved PIV to study unsteady aerodynamics. *IEEE Trans. Instr. Meas.* **59**, 3262-3269.

33. Echols WH, Young JA. 1963 Studies of portable air-operated aerosol generators. *No. NRL-5929*, Naval Research Laboratory Washington, DC.

34. Gurka R, Krishnanmoorthy K, Ben-Gida H, Kircehhfer AJ, Kopp GA, Guglielmo C. G. 2017 Flow pattern similarities in the near wake of three bird species suggest a common role for unsteady aerodynamic effects in lift generation. *The Royal Soc.: Interface Focus* **7**, 20160090.

35. Raffel M, Willert C, Wereley S. Kompenhans J. 2007 *Particle Image Velocimetry: A Practical Guide*, 2nd Ed. Springer.

36. Anderson JM, Streitlien K, Barret DS, Triantafyllou MS. 1998 Oscillating airfoils of high propulsive efficiency. *J. Fluid Mech.* **360**, 41-72.

37. Shyy W, Lian Y, Tang J, Viieru D, Liu H. 2008 *Aerodynamics of Low Reynolds Number Flyers*. UK: Cambridge University Press.

38. Rosti ME, Kamps L, Bruecker C, Omidyeganeh M, Pinelli A. 2017 The PELskin project-part V: towards the control of the flow around aerofoils at high angle of attack using a self-activated deployable flap. *Meccanica* **52**(8), 1811-1824.

39. Spedding GR, Rosén M, Hedenström A. 2003 A family of vortex wakes generated by a thrush nightingale in free flight in a wind tunnel over its entire natural range of flight speeds. *J. Exp. Bio.* **206**, 2313-2344.

40. Taylor GI. 1938 The spectrum of turbulence. *Proc. Royal Soc. London A: Math. Phys. Eng. Sci.* **164**(919), 476-490.

41. Chin DD, Lentink D. 2016 Flapping wing aerodynamics: from insects to vertebrates. *J. Exp. Bio.* **219**(7), 920-932.

42. Smith MR, Donnelly RJ, Goldenfeld N, Vinen WF. 1993 Decay of vorticity in homogeneous turbulence. *Phys. Rev. Lett.* **71**(16), 2583.

43. Pope SB. 2000 *Turbulent Flows*. UK: Cambridge University Press.

44. Adrian RJ, Christensen KT, Liu ZC. 2000 Analysis and interpretation of instantaneous turbulent velocity fields. *Exp. fluids* **29**(3), 275-290.

45. Ohkitani K, Kishiba S. 1995 Nonlocal nature of vortex stretching in an inviscid fluid. *Phys. Fluids* **7**(2), 411-421 (1995).

46. Tsinober A. 2013 *The Essence of Turbulence as a Physical Phenomenon: With Emphasis on Issues of Paradigmatic Nature*, Springer Science and Business Media.

47. Nomura KK, Post GK. 1998 The structure and dynamics of vorticity and rate of strain in incompressible homogeneous turbulence, *J. Fluid Mech.* **377**, 65-97 (1998).

48. Bachmann T, Wagner H. 2011 The three dimensional shape of serrations at barn owl wings: towards a typical natural serration as a role model for biomimetic applications. *J. Anatomy* **219**, 192-202.

**Tables captions**

Table 1: Morphological characteristics of the birds flown along with the characteristic flow numbers for the experiments performed at AFAR.

Table 2: A summary of the datasets collected during the owl flight experiments.

Table 3: Averaged angle of attack (γ) for five different experiments. The angle is calculated when the wingtip is parallel to the body and no twist is observed over the wing.

Table 4: Morphological characteristics of the birds flown along with the experimental parameters used for the fluid dynamic comparison, similar units to table 1.

Table 5: Characteristic flow scales at the near wake based on the auto-correlation of the velocity fields. The scales are normalized by the respective wing chord length.

**Figures captions**

Figure 1: PIV and kinematic imaging fields of view (FOV). The locations of the measured FOVs are at the center of the tunnel, observing a streamwise-normal plane. The PIV FOV was 13x13cm$^2$ and the kinematic FOV was 52x52cm$^2$. The distance between the two FOV's was 18-19 cm.

Figure 2: Sequence of instantaneous images showing a full wingbeat cycle of the boobook owl as the owl moves from right to left as it did in the wind tunnel.

Figure 3: Wingbeat kinematics: non-dimensional amplitude of the wingtip versus time. The solid vertical lines illustrate the wings transition phases. The dashed line illustrates the points during the downstroke where the angle of attack was calculated.

Figure 4: Angle of attack geometrical location for the flapping wing, based on the wingtip and wind speeds as well as the wing position in respect to the body.

Figure 5: Near-wake flow features of the boobook owl while flying in a flapping mode. The owl flew from right to left. (i) Wake reconstruction - The wake was sampled behind the wing at different spanwise sections: a) between the primary and secondary remiges; experiment #9, b) outer region of the wing; the furthest location in the primary remiges; experiment #8 and (c) middle of the primary remiges; experiment #3. Contours represent the values of spanwise vorticity and the vectors depict the two-dimensional, two-component velocity field in the near wake. (ii) Wingtip displacement - The wingtip displacement is plotted against downstream chord length to directly correlate with the respective wake. The vertical black lines in each graph represent the transition from upstroke to downstroke or downstroke to upstroke respectively.

Figure 6: Distribution of concentrated spanwise vorticity regions at the wake of the three birds. The histogram is based on blob analysis performed on the wake reconstruction contours appear in figure 5a for the owl and in figure 4 and 5 in Gurka et al. (2017) for the sandpiper and starling, respectively. The top figures illustrate the positive spanwise vorticity selections and the bottom showing the negative ones. The left figures compare the owl with the sandpiper, the middle compare it with the starling and the right figures compares the sandpiper with the starling.

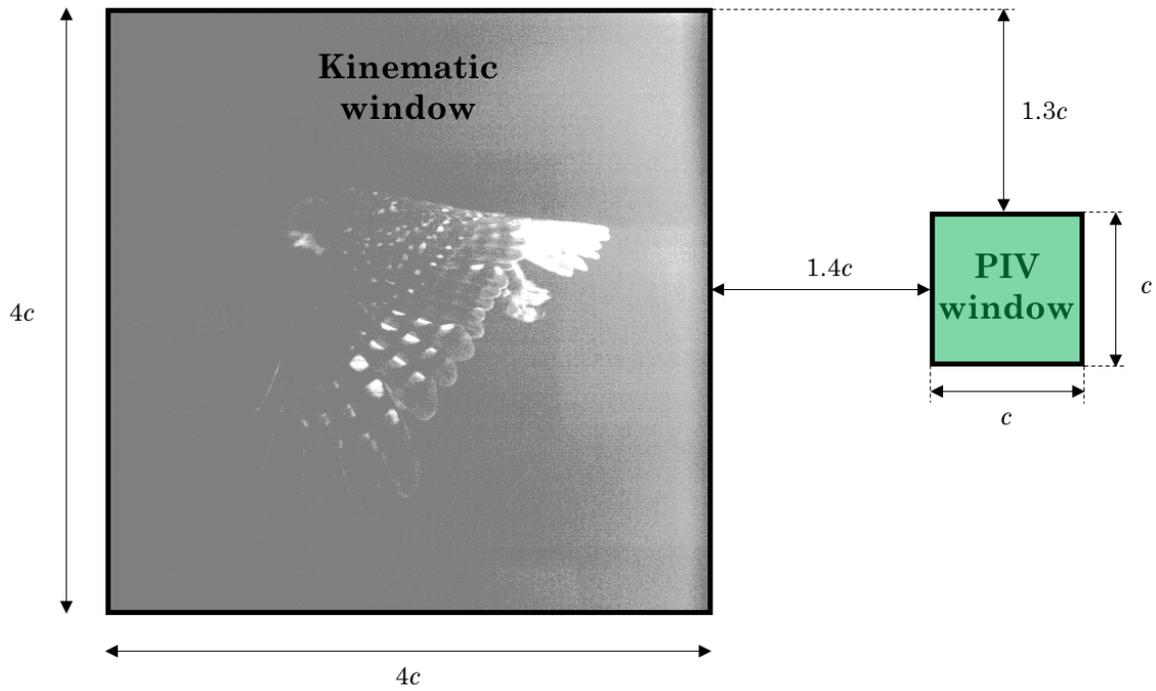

**Figure 1:** PIV and kinematic imaging fields of view (FOV). The locations of the measured FOVs are at the center of the tunnel, observing a streamwise-normal plane. The PIV FOV was 13x13cm$^2$ and the kinematic FOV was 52x52cm$^2$. The distance between the two FOV's was 18-19 cm.

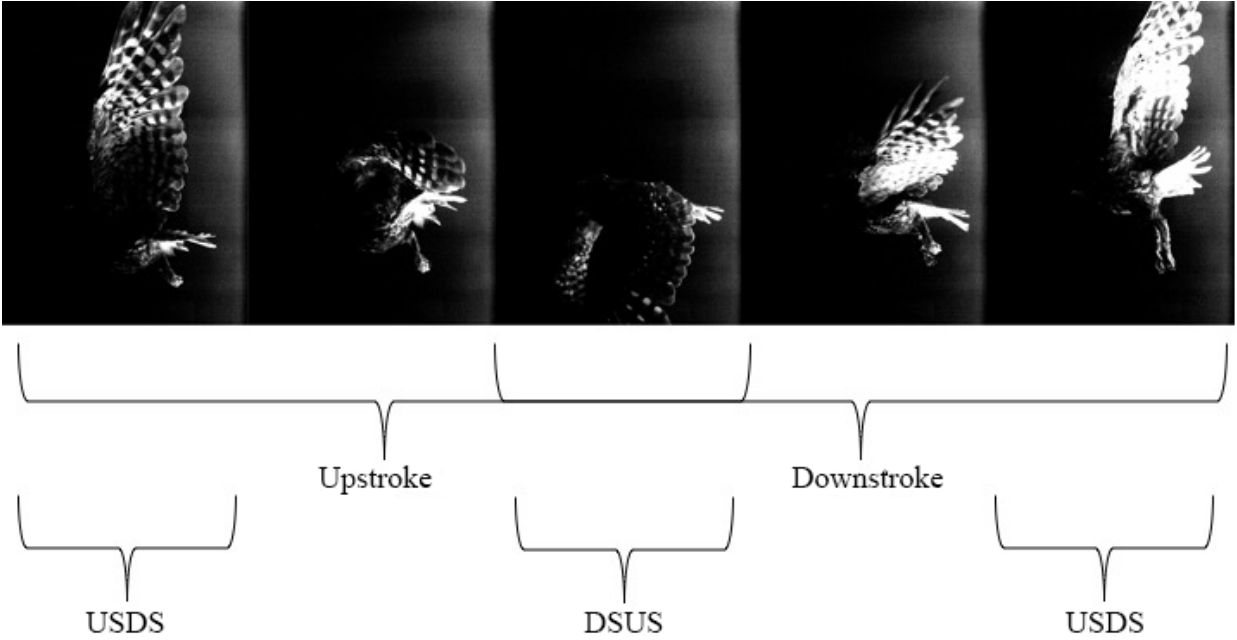

**Figure 2**: Sequence of instantaneous images showing a full wingbeat cycle of the boobook owl as the owl moves from right to left as it did in the wind tunnel.

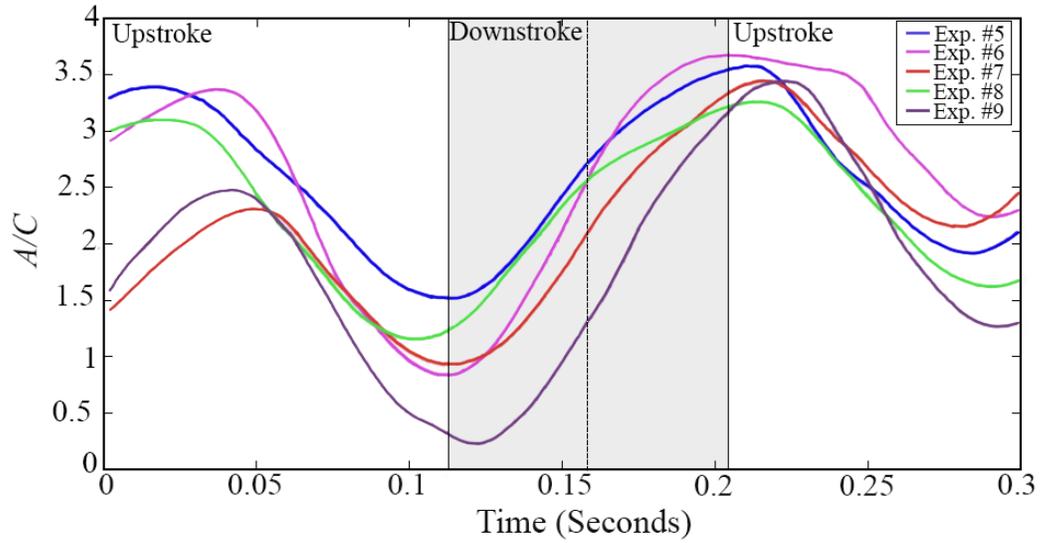

**Figure 3:** Wingbeat kinematics: non-dimensional amplitude of the wingtip versus time. The solid vertical lines illustrate the wings transition phases. The dashed line illustrates the points during the downstroke where the angle of attack was calculated.

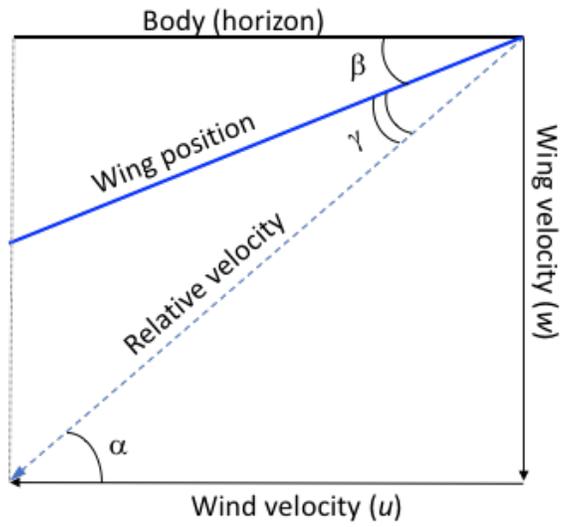

**Figure 4**: Angle of attack geometrical location for the flapping wing, based on the wingtip and wind speeds as well as the wing position in respect to the body.

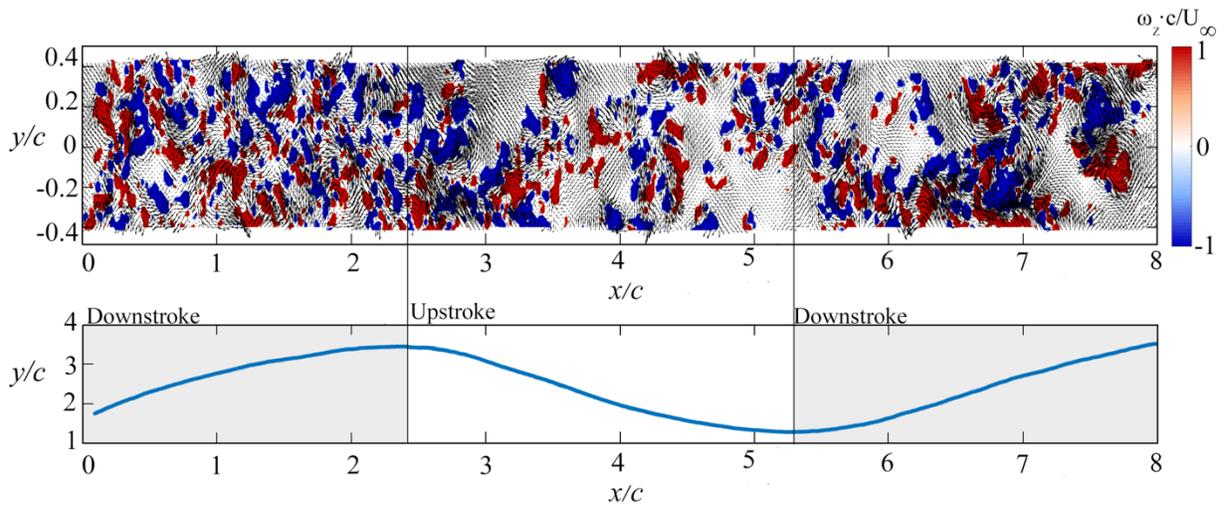

a) Exp#9

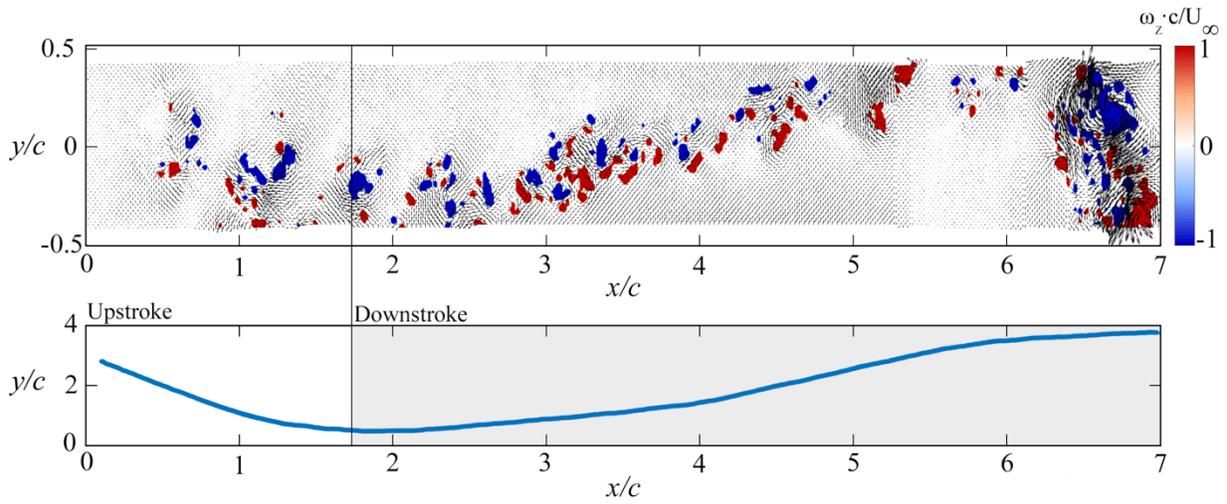

b) Exp#8

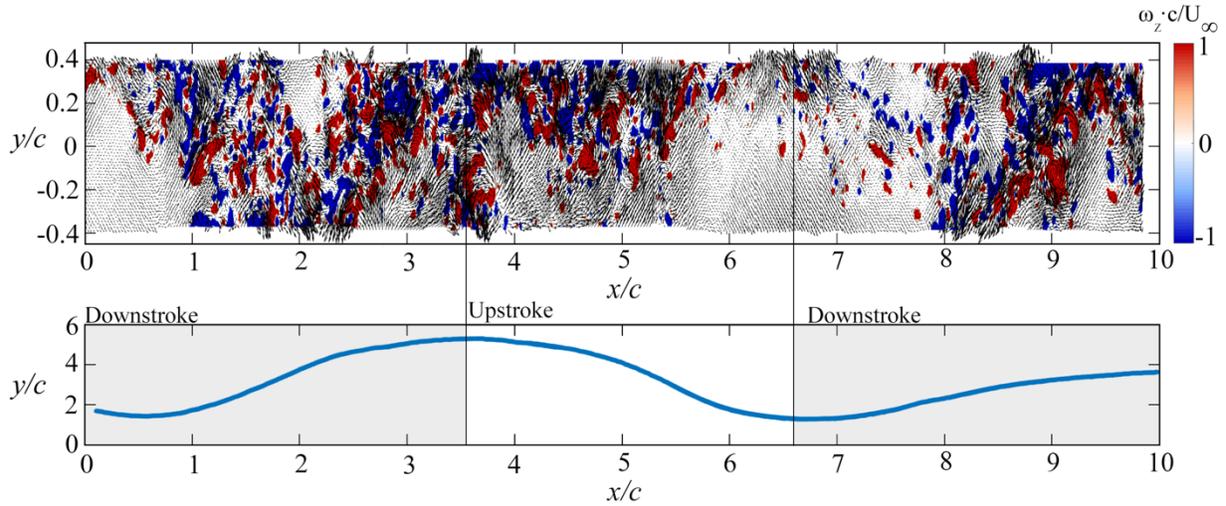

c) Exp#3

**Figure 5**: Near-wake flow features of the boobook owl while flying in a flapping mode. The owl flew from right to left. (i) Wake reconstruction - The wake was sampled behind the wing at different spanwise sections: a) between the primary and secondary remiges; experiment #9, b) outer region of the wing; the furthest location in the primary remiges; experiment #8 and (c) middle of the primary remiges; experiment #3. Contours represent the values of spanwise vorticity and the vectors depict the two-dimensional, two-component velocity field in the near wake. (ii) Wingtip displacement - The wingtip displacement is plotted against downstream chord length to directly correlate with the respective wake. The vertical black lines in each graph represent the transition from upstroke to downstroke or downstroke to upstroke respectively.

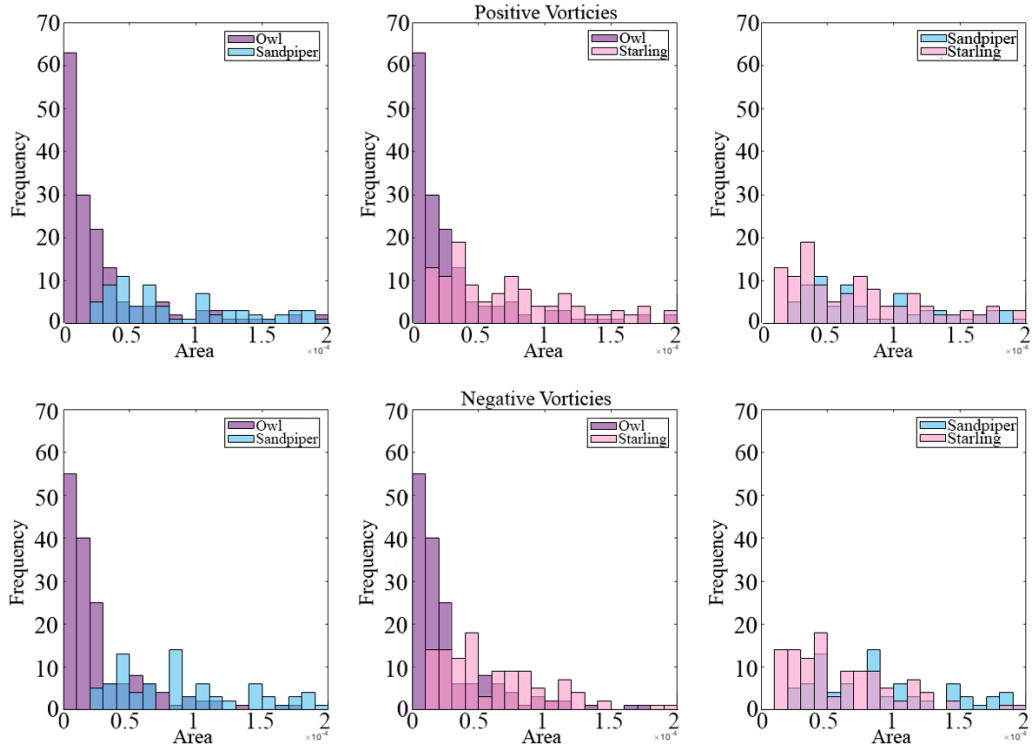

**Figure 6:** Distribution of concentrated spanwise vorticity regions at the wake of the three birds. The histogram is based on blob analysis performed on the wake reconstruction contours appear in figure 5a for the owl and in figure 4 and 5 in Gurka et al. (2017) for the sandpiper and starling, respectively. The top figures illustrate the positive spanwise vorticity selections and the bottom showing the negative ones. The left figures compare the owl with the sandpiper, the middle compare it with the starling and the right figures compares the sandpiper with the starling.

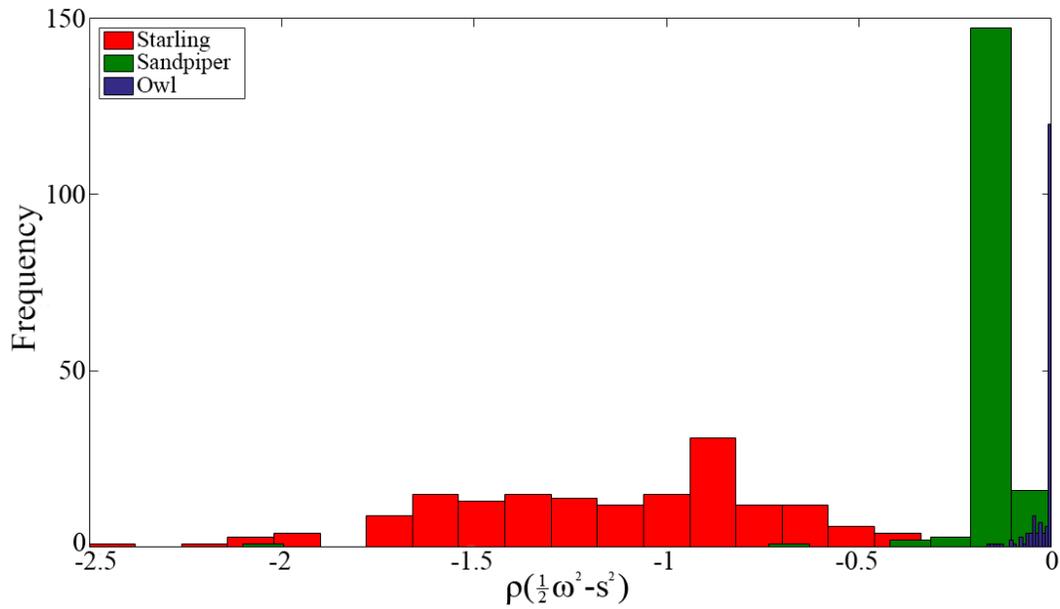

**Figure 7**: Distribution of the right-hand side term in equation 1; pressure Hessian at the near wake region for the three birds. The histogram is based on calculating the vorticity and the strain fields for experiment #9 for the owl and experiential data for the sandpiper and starling were deduced from Gurka et al. (2017). Blue, red and green histograms correspond to the owl, starling and sandpiper wake data, respectively.